\documentclass[a4paper,11pt]{article}
\usepackage{pos}

\title{Improved Gamma-Hadron Separation\\ for PeV Photon Searches with IceCube}
\ShortTitle{IceCube Gamma-Hadron Separation}

\author{The IceCube Collaboration \\{\normalsize \normalfont(a complete list of authors can be found at the end of the proceedings)}\\}

\emailAdd{fgs@udel.edu}
\emailAdd{federico.bontempo@kit.edu}

\abstract{IceTop, the km$^2$ surface array of the IceCube Neutrino Observatory at the South Pole, is sensitive to air showers of all primary particles, including gamma rays. In particular, in the PeV energy range, the combination of IceTop and IceCube’s deep optical detector provides excellent gamma-hadron separation. 
Almost all air showers induced by cosmic-ray protons and heavier nuclei in this energy range contain high-energy muons detectable by the deep detector, while most photon-induced air showers do not. 
Therefore, IceCube’s deep detector can be used to strongly suppress hadronic background in photon searches. Furthermore, the lateral distribution of the air-shower signal in IceTop provides additional gamma-hadron separation. 
In the PeV energy range, the gamma-hadron separation achieved is better than $10^{-3}$: air-shower events measured with IceCube are suppressed more than $1000$ times stronger than photon-induced showers of the same energy simulated with Sibyll 2.3d. 
This improved gamma-hadron separation in combination with an extension of the energy range to lower energies provides discovery potential for future searches for PeV photon sources in IceCube’s field of view.

\vspace{4mm}

{\bfseries Corresponding authors:}
Frank G.~Schroeder$^{1,2*}$, 
Federico Bontempo$^{2}$\\
{$^{1}$ \itshape Bartol Research Institute, Department of Physics and Astronomy, University of Delaware}\\
{$^{2}$ \itshape Institute for Astroparticle Physics, Karlsruhe Institute of Technology (KIT)}\\[4mm]
$^*$ Presenter
}

\ConferenceLogo{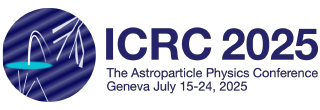}

\FullConference{39th International Cosmic Ray Conference (ICRC2025)\\
 15–24 July 2025\\
Geneva, Switzerland\\}

\begin{document}
\maketitle

\section{Introduction}
The IceCube Neutrino Observatory at the South Pole is a unique experiment consisting of the km$^2$-size IceTop surface array for air showers above a km$^3$ optical array deep in the Antarctic ice.
The latter detects not only particles from neutrino-induced interactions, but mostly high-energy atmospheric muons from cosmic-ray air showers.
Because high-energy muons are absent in most gamma-ray induced air showers up to several PeV of energy, the combination of IceTop and the deep array can be used to separate hadronic cosmic rays from gamma rays.
Earlier IceCube studies focused on the search for multi-PeV gamma rays~\cite{IceCube:2019scr} and were performed before LHAASO detected first sources of PeV photons with a surface array of similar size as IceTop~\cite{LHAASO:2021gok}. 
Although LHAASO has a different field-of-view (FOV) than IceCube for gamma rays, we expect that photon-induced air showers are already present in IceCube's data if any similar source is in IceCube's FOV.

\begin{figure}[t]
\begin{center}
    \includegraphics[width=0.7\linewidth]{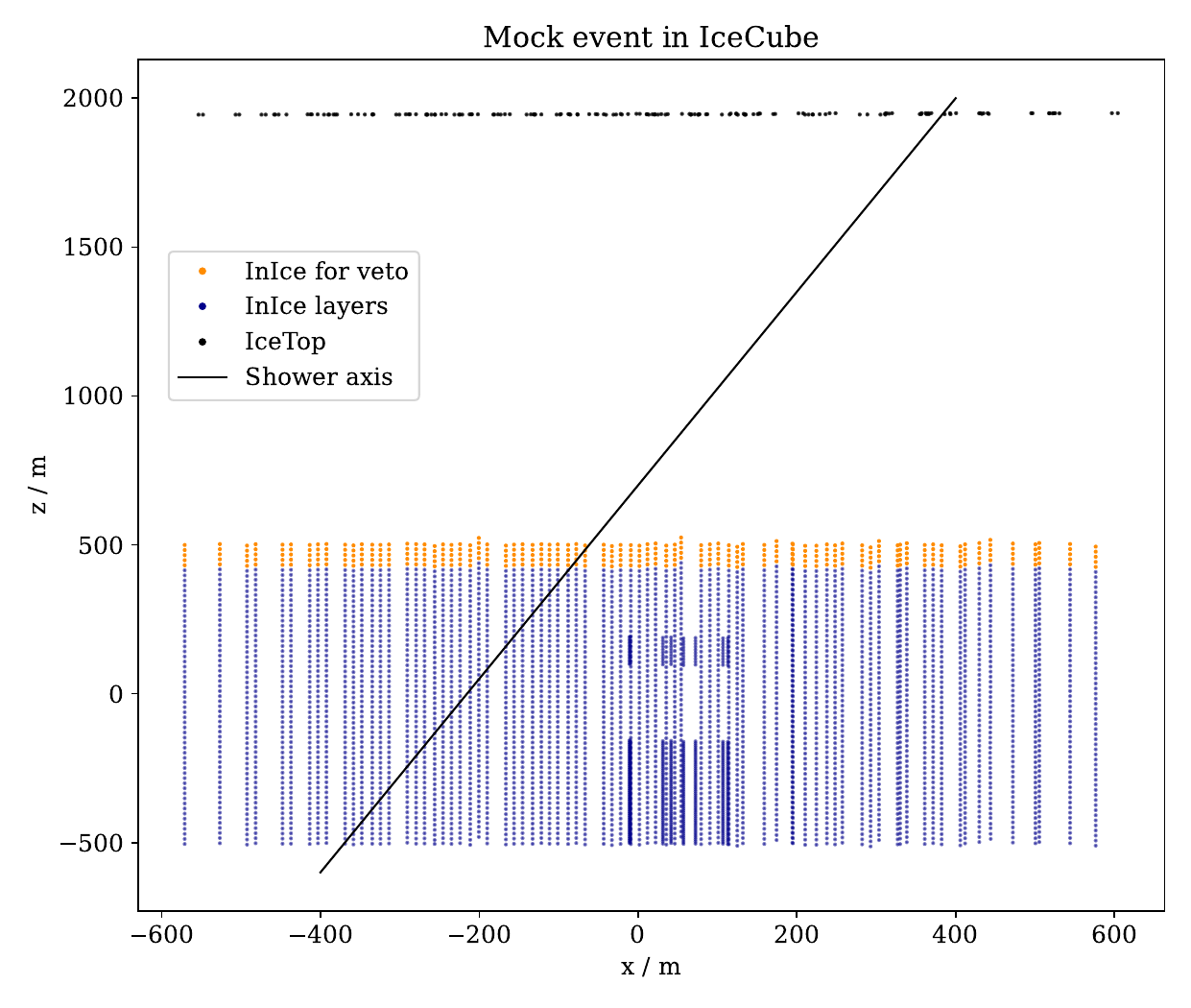}
    \caption{Projection of the IceCube detector with its IceTop surface array on top and the deep optical in-ice array. For air showers whose axis intersects both arrays, the in-ice array is used as a veto, rejecting air showers that have any signal in the top five layers.
    }
    \label{fig_ArrayVeto}
\end{center}
\end{figure}

Because the flux of LHAASO's PeV sources is much stronger at energies below a PeV, we have improved the reconstruction methods for air showers to be more efficient in the sub-PeV range and improved the method of gamma-hadron separation to provide a higher rejection of hadronic cosmic rays~\cite{BontempoThesis}.
Here, we summarize the method of gamma-hadron-separation: the top layers of the deep array serve as a veto to reject hadronic showers (see Fig.~\ref{fig_ArrayVeto}), and the compactness of the lateral distribution, measured by IceTop through a quantity named Charge Distance (CD), provides additional separation power.
Finally, we estimate the number of events and significance expected for potential LHAASO-like sources, which illustrates IceCube's discovery potential for PeV gamma-ray sources. 


\begin{figure}[t]
\begin{center}
    \includegraphics[width=0.49\linewidth]{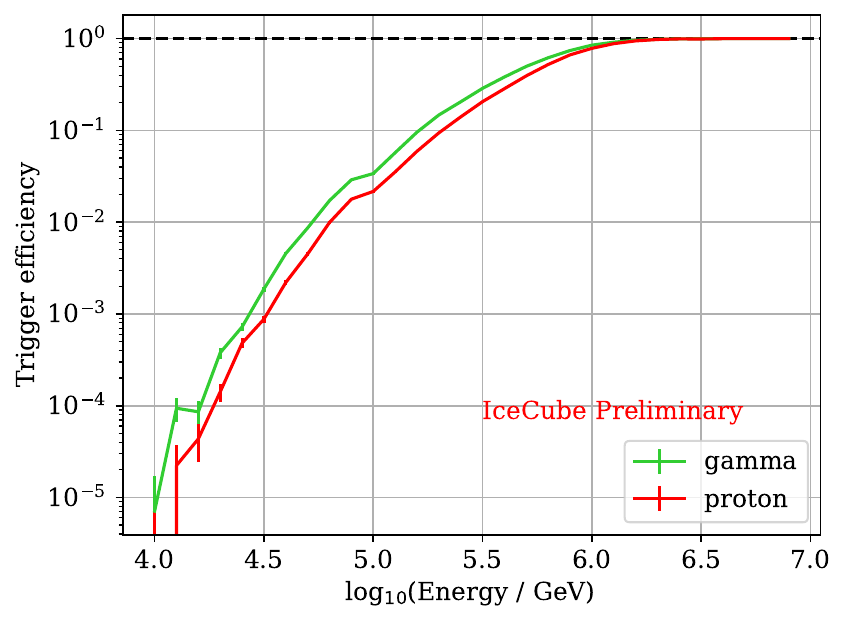}
    \hfill
    \includegraphics[width=0.49\linewidth]{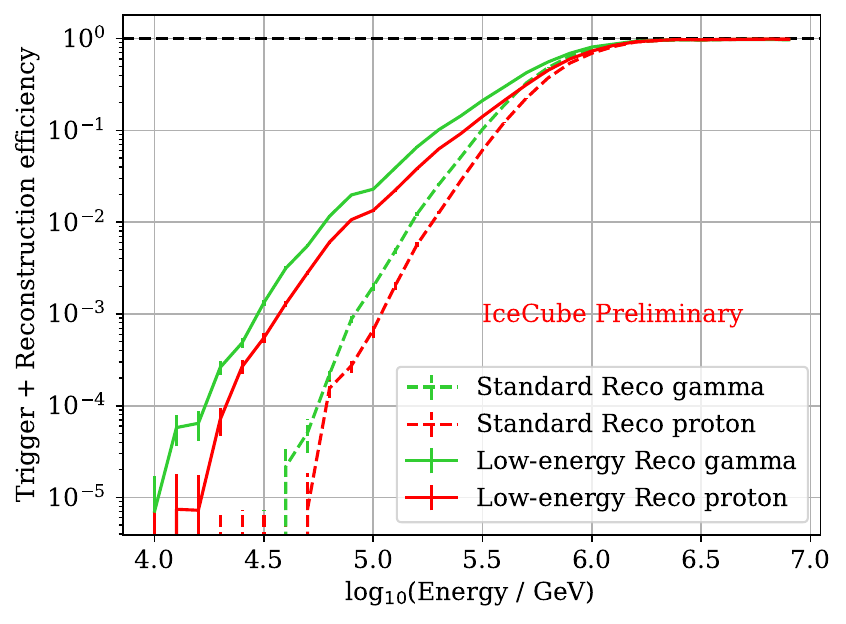}
    \caption{Detection efficiency of IceTop for protons and photons (left) and combined detection and reconstruction efficiency (right) over energy of the primary particle. The latter requires a successful reconstruction of the shower direction and energy proxy $S_\mathrm{125}$ within certain quality cuts (see Ref.~\cite{BontempoThesis} for details).}
    \label{fig_efficiency}
\end{center}
\end{figure}

\begin{figure}[t]
\begin{center}
    \includegraphics[width=0.49\linewidth]{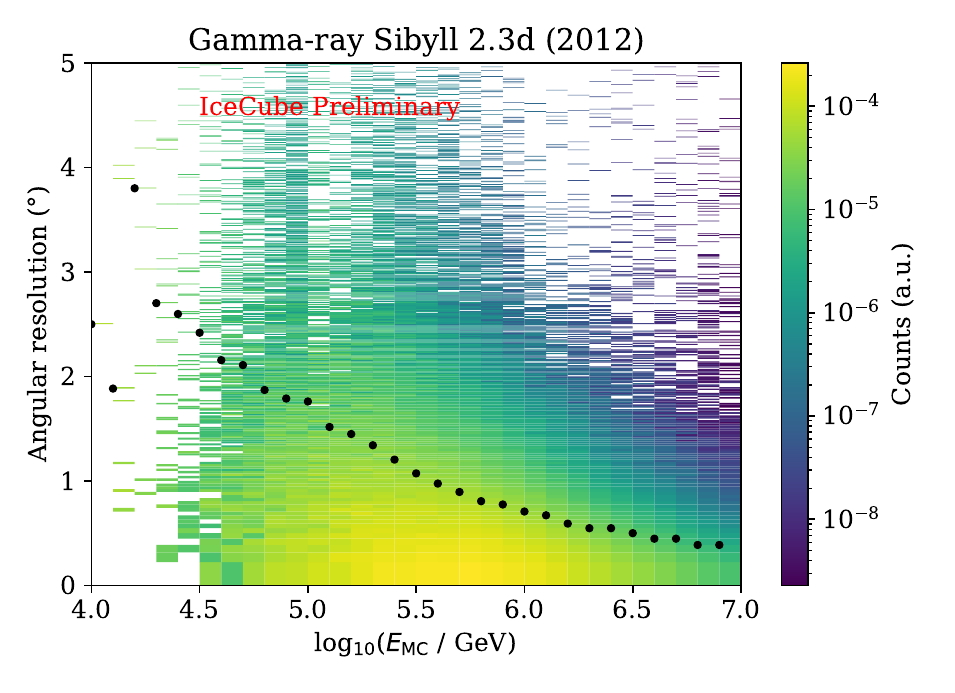}
    \hfill
    \includegraphics[width=0.49\linewidth]{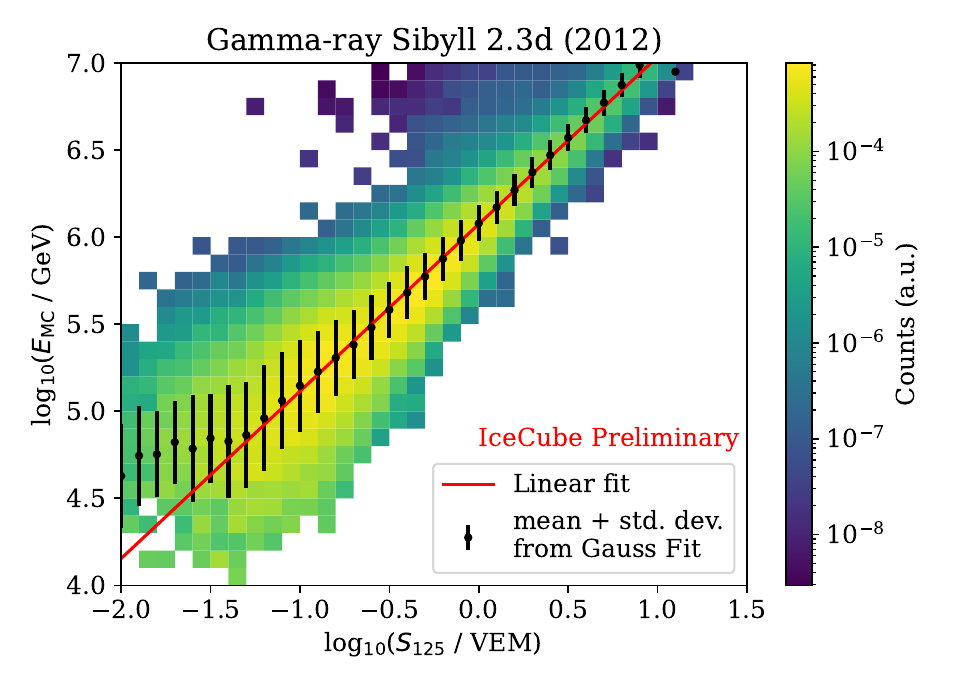}
    \includegraphics[width=0.49\linewidth]{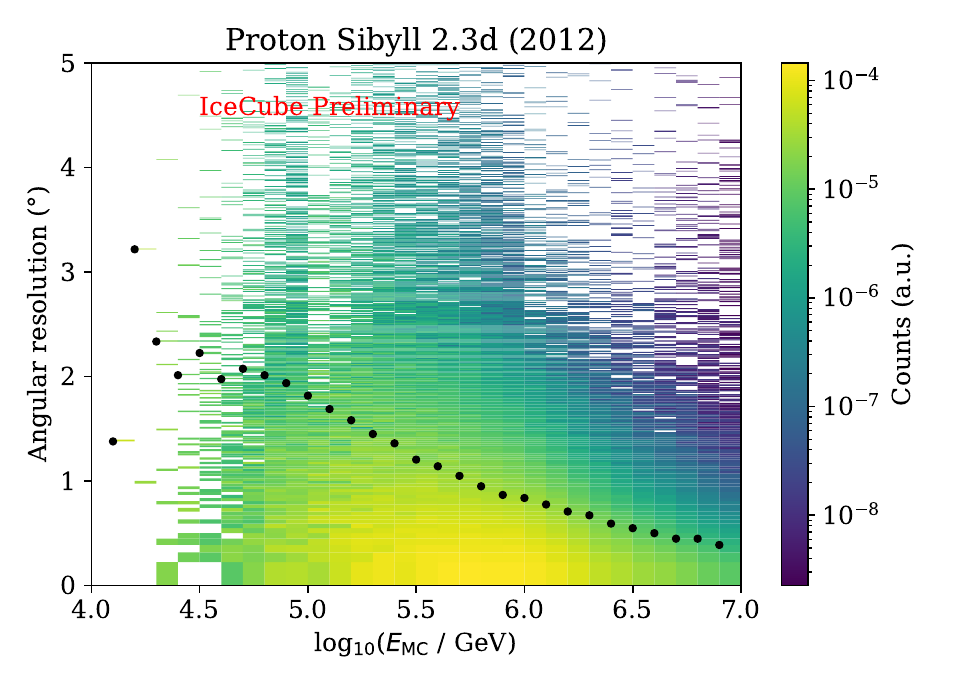}
    \hfill
    \includegraphics[width=0.49\linewidth]{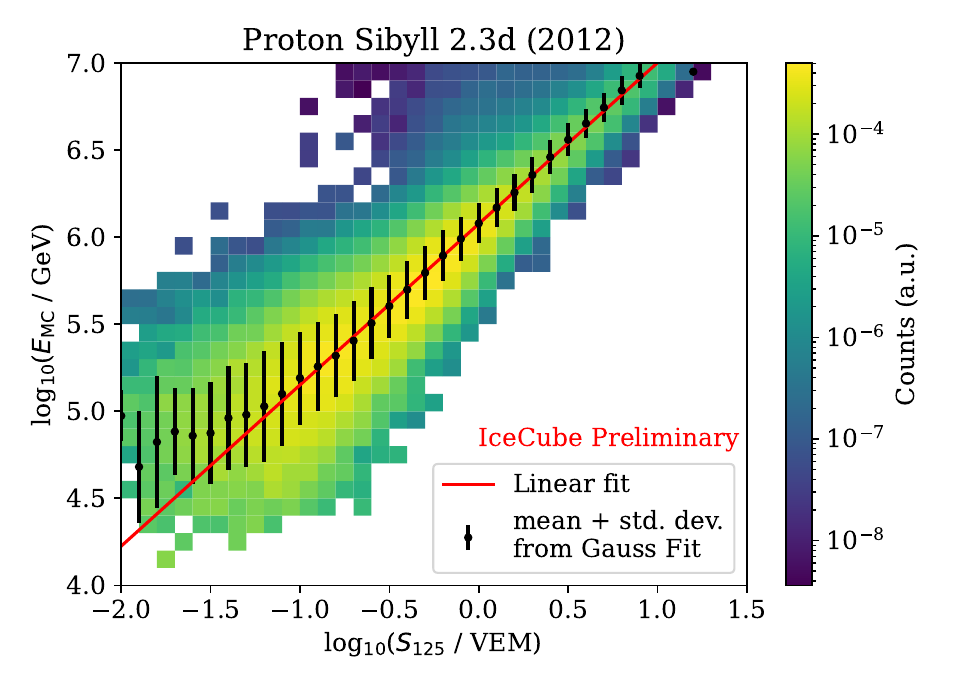}
    \caption{Left: Angular resolution of the low-energy IceTop reconstruction, determined as the $68\,\%$ percentile $\Delta \psi$ of the deviation between the true and reconstructed shower axis for simulated air showers (shown as black dots). Right: The energy proxy $S_{125}$, the signal strengths in IceTop at $125\,$m distance from the shower axis, is approximately linearly correlated with the energy of the primary particle. Photon (top) and proton (bottom) simulations show a similar angular resolution and energy correlation with $S_{125}$.}
    \label{fig_resolutions}
\end{center}
\end{figure}

\begin{figure}[t]
\begin{center}
    \includegraphics[width=0.8\linewidth]{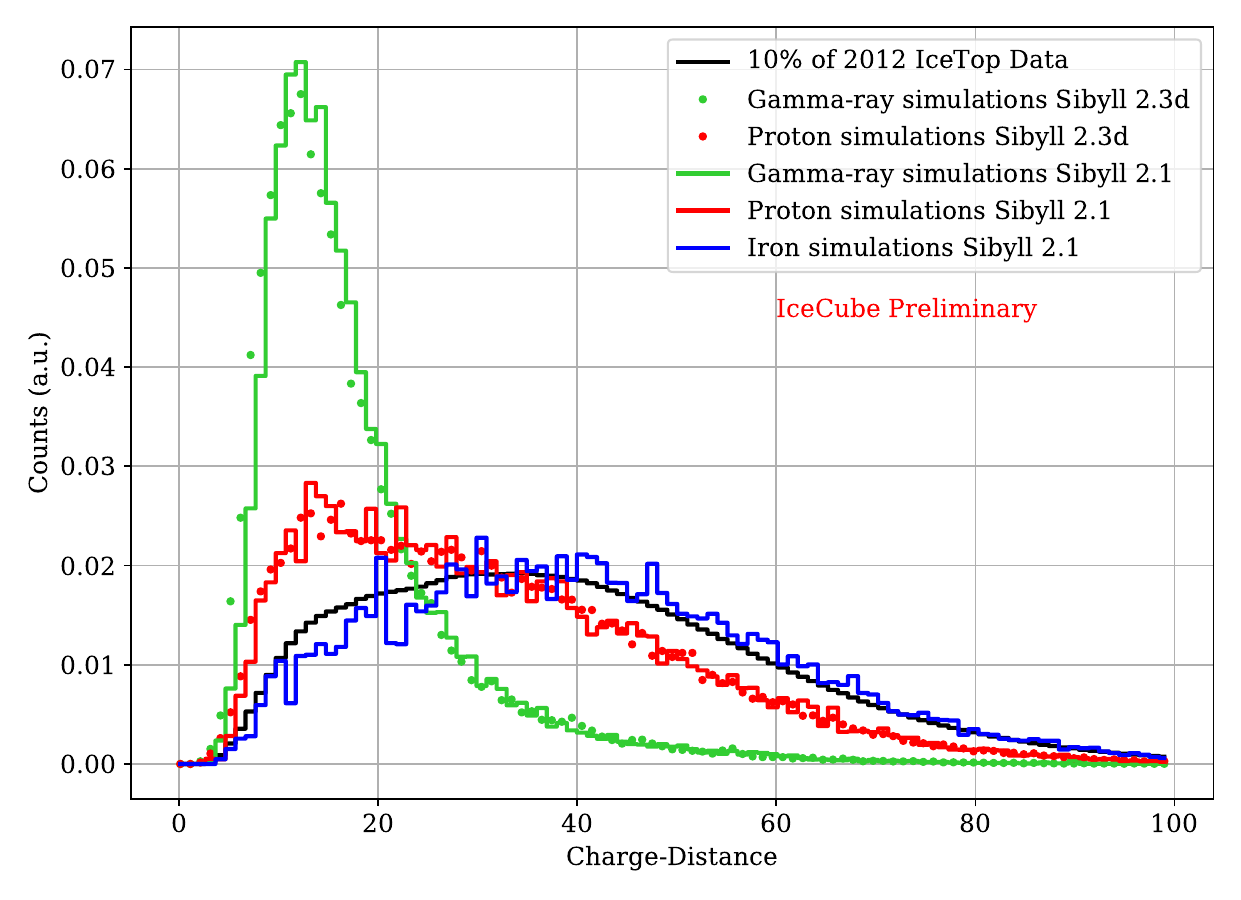}
    \caption{Distribution of the Charge Distance discriminator based on the lateral distribution of IceTop hits of an air shower for simulations and measured data of the 2012 run.
    }
    \label{fig_ChargeDistance}
\end{center}
\end{figure}

\begin{figure}[t]
\begin{center}
    \includegraphics[width=0.9\linewidth]{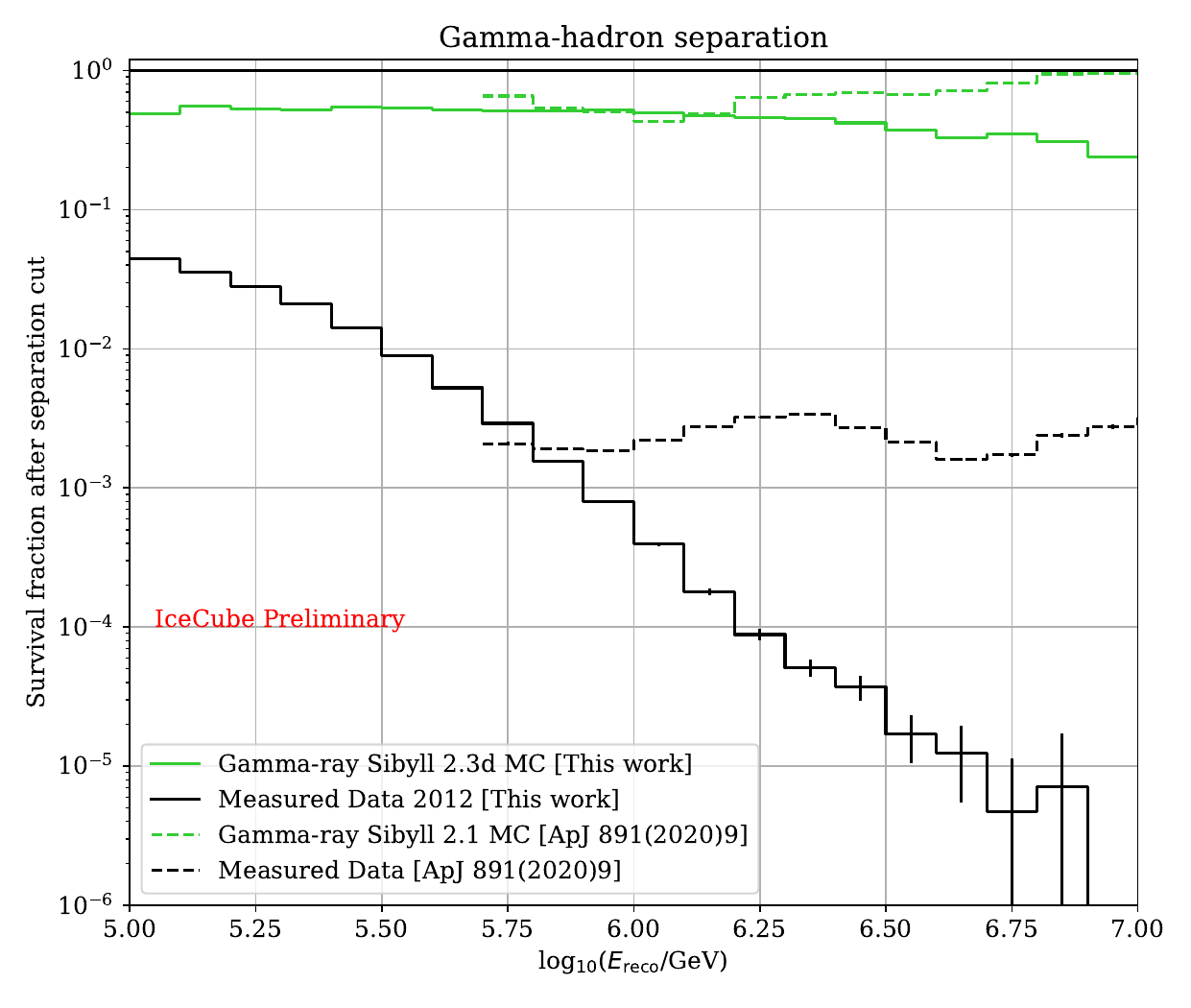}
    \caption{Gamma-hadron separation achieved by the combination of the (absence of) in-ice signal and the Charge Distance measurement of IceTop. Shown is the passing fraction of photon simulations and measured data of the 2012 run, where the measured data is dominated by hadronic cosmic rays. In comparison with IceCube's previous gamma-ray search~\cite{IceCube:2019scr}, the separation power extends to lower energies and is significantly enhanced in the PeV energy range.
    }
    \label{fig_gammaHadronSeparation}
\end{center}
\end{figure}

\section{Method}
To study gamma-hadron separation with IceCube, we have used simulations and measured air-shower data; the latter are predominantly from hadronic cosmic rays.
For the simulations, we have produced several $10,000$'s of CORSIKA air-shower events (at least $1,000$ per energy bin of $\Delta \log_{10} E = 0.1$) using the interaction model Sibyll 2.3d, and also used existing simulation sets with Sibyll 2.1, as we have observed no relevant difference regarding the air-shower parameters studied in this work for gamma-hadron separation.
Nonetheless, we also use measured data to study the hadronic background for two reasons:
First, it is known that simulations do not fully reproduce measurements~\cite{Verpoest:2025gui}; second, because the simulations are computationally expensive, measured events provide a much higher statistics in the relevant energy range of $E \lesssim 10\,$PeV. 
Therefore, we use approximately one year of IceCube events: the 2012 run recorded between April 2012 and May 2013. 
To keep the analysis blinded for future point-source searches, we have randomized the right ascension of the measured air showers.

As first major improvement over previous analyses, we have lowered the energy threshold by improving the reconstruction efficiency towards lower energies (see Fig.~\ref{fig_efficiency}), at which IceTop is only partially efficient. 
IceTop, built of 81 pairs of ice-Cherenkov tanks~\cite{IceCube:2012nn}, triggers upon coincident signals in three pairs of tanks (so 6 tanks total).
Nonetheless, many analyses only consider events with at least five pairs of tanks with signals (10 tanks total), because of the higher precision for the air-shower reconstruction.
Using the established functions describing the lateral distribution and arrival time of the shower front recorded by IceTop, we have adapted the fitting procedure to feature fewer free parameters for air-showers with signal in at least three pairs of IceTop tanks, and still keep a decent angular resolution and energy reconstruction (see~Fig.~\ref{fig_resolutions}).

The reconstruction of the air-shower arrival direction and energy is based only on signals recorded coincidentally in both tanks of a pair to reduce the impact of random background. 
However, signals in only one of the two tanks are often caused by muons and can be used for gamma-hadron separation.
Therefore, a calibration method for these single-tank signals was developed (for details see Ref.~\cite{BontempoThesis}), and they are included in this analysis for the calculation of a quantity called Charge Distance ($CD$), which measures the compactness of the lateral distribution of an air shower recorded by IceTop:

\begin{equation}
CD = \log_{10} \sum_i \left(\frac{q_i}{0.1\,\mathrm{VEM}}\right)^{d_i / 10\,\mathrm{m}},   
\end{equation}
with $q_i$ the signal size 'charge' in the $i^\mathrm{th}$ tank measured in units of vertical equivalent muons (VEM) and $d_i$ the distance of that tank from the shower core at ground in meters; the constants before the units were chosen ad-hoc after testing constants of different orders of magnitude.
Because photon-induced air showers have a more compact footprint, their $CD$ is on average smaller than those of hadronic air showers, as seen in Fig.~\ref{fig_ChargeDistance}.

Due to the known deficiencies of hadronic interaction models, the $CD$ distribution should not be used to determine the mass composition without further studies, but the proton and iron simulations show that the simulated $CD$ is at least roughly consistent with IceTop's measurement. 
We used the photon simulations and measured air showers to decide on cuts for the gamma-hadron simulations, where the measurements are dominated by hadronic showers.

The two cuts providing gamma-hadron separation, i.e., the number of in-ice layers used as a veto and the percentile of the $CD$ cut, were optimized to provide maximum significance for LHAASO-like mock sources with one year of 2012 IceCube data (see next section.)
The number of layers needs to be optimized for two reasons: some high-energy muons do not cause a detectable signal in every layer they pass, so more veto layers provide a better rejection; however, more veto layers also increase the chance of accidentally vetoing a gamma-ray shower because of random background signals.
Similarly, the $CD$ percentile used as a cut needs to  balance rejecting hadronic cosmic rays versus keeping gamma-ray candidates.
As the average $CD$ increases with energy, also the cut used on the $CD$ is energy dependent and has been tuned as a function of IceTop's energy proxy $S_\mathrm{125}$~\cite{IceCube:2012nn} to keep $85\,\%$ of the photons at each energy:
\begin{equation}
CD_\mathrm{cut} = 21.7 \cdot \log_{10}(S_\mathrm{125}/\mathrm{VEM})+35.1.
\end{equation}

As a result of the optimization, we reject all air showers that have a signal in the five top layers of the in-ice detector or have a $CD$ larger than the $85^\mathrm{th}$ percentile of the CD distribution for photons at the respective shower energy.

\section{Results}
Using the top five layers of the in-ice detector as veto and the $CD$ cut at IceTop, together, provides gamma-hadron separation by three to four orders of magnitude in the PeV energy range.
Because the fraction of hadronic showers without high-energy muons decreases with energy, the separation power strongly increases with energy, suppressing hadronic showers relative to photon showers by one order of magnitude around $100\,$TeV and more than four orders of magnitude above $3\,$PeV.
The passing fractions for photon simulations and measured data can be seen in Fig.~\ref{fig_gammaHadronSeparation} over the energy reconstructed through the correlation shown in Fig.~\ref{fig_resolutions}.
As the measured data are dominated by hadronic cosmic rays, the ratio between the survival fraction of the photon simulations and the survival fraction of the measured data provides a measure for the gamma-hadron separation: the gamma-hadron separation improves with energy from about one order of magnitude around $0.1\,$PeV to more than four orders of magnitude above $3\,$PeV.


The unwanted rejection of about half of the photons is mostly due to accidental background signal in the upper layers of the in-ice detector, as by construction only $15\,\%$ of photons are rejected due to the $CD$ cut. 
Only at higher energies of several PeV, the passing fraction of photons decreases further, because a significant fraction of photon-induced showers contain at least one high-energy muon producing an in-ice signal.
The previous search had a much higher passing fraction for photons, by tolerating some in-ice signal coincident with an air shower measured with IceTop~\cite{IceCube:2019scr}. 
However, not tolerating any in-ice signal at all provides much better gamma-hadron separation without decreasing the survival fraction of photon candidates in the energy range around $1\,$PeV and, hence, is preferred to search for gamma-ray sources emitting photons up to a maximum energy in that range. 

\begin{figure}[t]
\begin{center}
    \includegraphics[height=4.9cm]{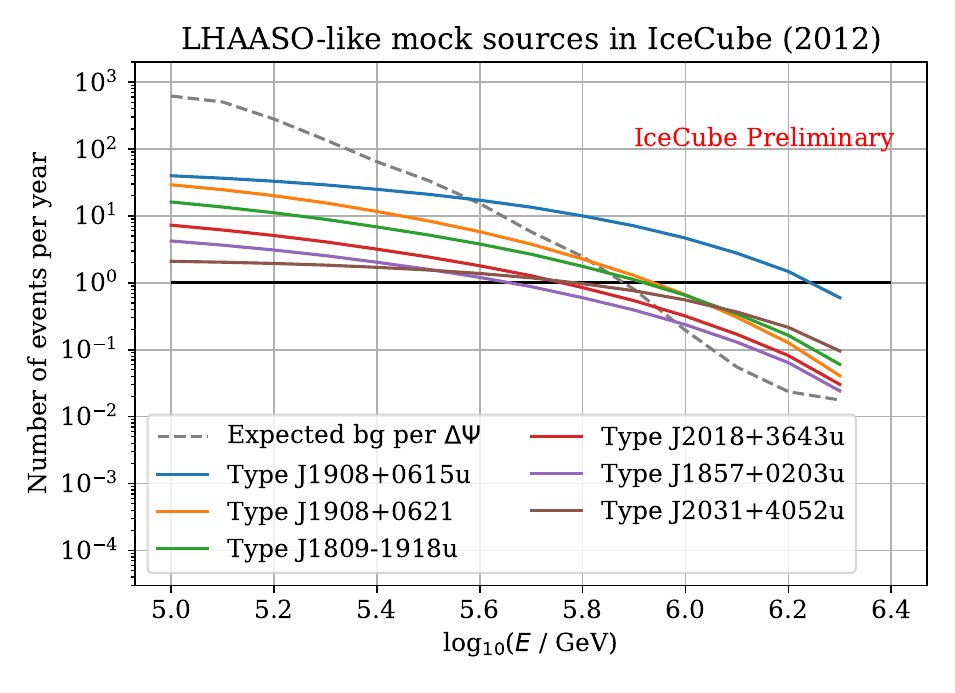}
    \hfill
    \includegraphics[height=4.9cm]{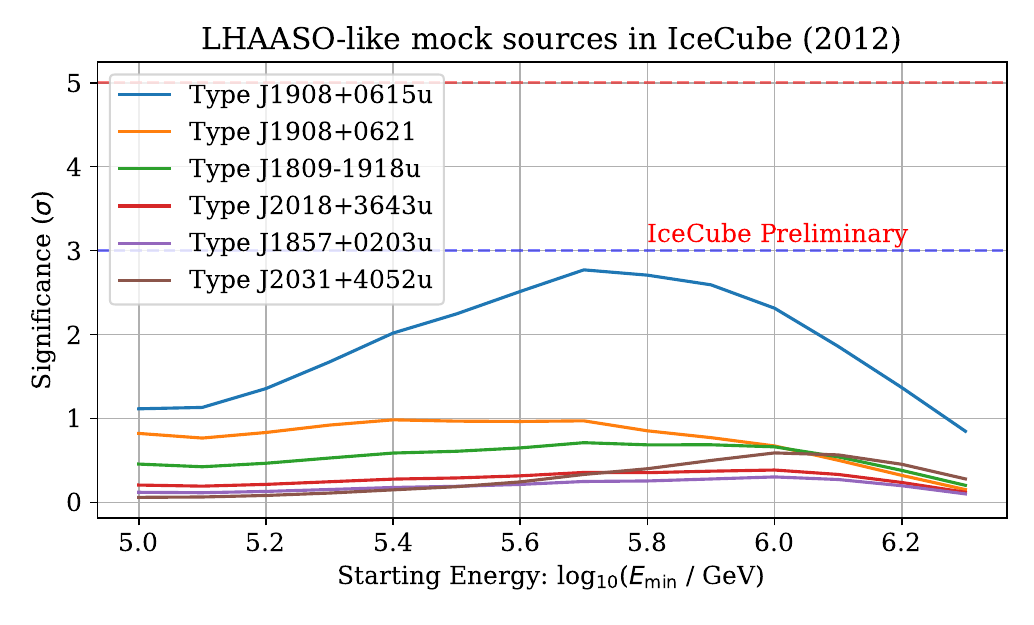}
    \caption{Number of events (left) and statistical significance (right) that would be expected with one year of IceCube data for a source at $85^\circ$ elevation above IceCube with the same photon flux as some of the sources detected by LHAASO (none of the LHAASO sources is in IceCube's FOV, as LHAASO observes a different part of the sky). The calculation considers all expected signal and measured background events above the energy shown on the x-axis and within IceCube's angular resolution for gamma rays. }
    \label{fig_MockSourceDetection}
\end{center}
\end{figure}

To study the detection potential of gamma-ray sources with the new method of gamma-hadron separation, we have calculated the number of photons we would expect to arrive from hypothetical mock sources at $\theta = 5^\circ$ zenith angle above IceCube, assuming a flux corresponding to sources detected by LHAASO~\cite{LHAASO:2023gne}.
Taking as search region for a specific source the solid angle defined by the angular resolution ($\pm \Delta \psi$), we count the photon-induced air showers contained in IceTop.
Then, we multiply the number of these contained showers with the detection and reconstruction efficiencies and the passing fraction of the gamma-hadron separation to calculate the number of detected photons for each mock source (Fig.~\ref{fig_MockSourceDetection}, left).
Calculating the number of cosmic-ray background events passing the gamma-hadron selection from the same solid angle, we then derive the Li-Ma significance expected for each mock source as a function of the lowest energy considered in the analysis (Fig.~\ref{fig_MockSourceDetection}, right). 
We find that the strongest source would be detectable almost at 3 sigma in only one year of IceCube data with a starting energy of the analysis around $0.5\,$PeV.

This suggests that IceCube could discover corresponding sources when combining several years of data. 
However, that combination is not trivial and subject to future work, as the rising snow level above IceTop impacts the threshold, reconstruction, and $CD$ cut.
As another complication, most sources are expected in the Galactic Plane, which is only visible at zenith angles $\theta \gtrsim 25^\circ$, but IceCube's exposure for photons shrinks with zenith angle due to the requirement of the shower axis intersecting both the surface and in-ice array: the maximum effective area with the current method of gamma-hadron separation is about $0.25\,$km$^2$ for vertical showers, which is about $40\,\%$ of IceTop's geometric area for the containment cuts used in this work, but shrinks to 0 at a zenith angle of about $\theta_\mathrm{max}=33^\circ$.
Future work will thus aim at retaining more photons without reducing the gamma-hadron separation power, e.g., by improving the selection of in-ice signals contributing to the veto or by better discriminators from IceTop's measurements for those air showers not rejected by the in-ice detector.

\section{Conclusion}
IceTop, IceCube's surface array for air showers, serves not only as a cosmic-ray detector and as a veto of atmospheric background for neutrino observations with IceCube's in-ice detector.
It also serves as a photon detector, when reversing the roles and using the in-ice detector as a veto instead, to suppress hadronic air showers against photon-induced air showers.
Recent progress in analysis techniques lowered the detection threshold for such gamma-ray searches into the sub-PeV energy range and improved the gamma-hadron separation significantly. 
With both improvements combined, analyses of IceCube's existing data now have discovery potential for gamma-ray sources with a similar flux as the brightest PeV gamma-ray sources detected by LHAASO, if such a source existed in the sky close to the celestial South Pole.
To maximize the discovery potential, we plan to work on further improvements in the gamma-hadron separation and the combination of several years of data with different snow levels, before unblinding.
Also, IceCube's FOV contains only a small part of the Galactic Plane with limited exposure.
This shortcoming will be significantly improved by IceCube-Gen2~\cite{IceCubeGen2TDR}, which will not only increase the aperture by an order of magnitude, but also extend the FOV to include a significant fraction of the Galactic Plane~\cite{Schroeder:2024wgq}.

\bibliographystyle{ICRC}
\bibliography{bibliography}

\clearpage

\section*{Full Author List: IceCube Collaboration}

\scriptsize
\noindent
R. Abbasi$^{16}$,
M. Ackermann$^{63}$,
J. Adams$^{17}$,
S. K. Agarwalla$^{39,\: {\rm a}}$,
J. A. Aguilar$^{10}$,
M. Ahlers$^{21}$,
J.M. Alameddine$^{22}$,
S. Ali$^{35}$,
N. M. Amin$^{43}$,
K. Andeen$^{41}$
% ...
\\
\\
% ...
$^{10}$ Universit{\'e} Libre de Bruxelles, Science Faculty CP230, B-1050 Brussels, Belgium \\
% ...
$^{16}$ Department of Physics, Loyola University Chicago, Chicago, IL 60660, USA \\
$^{17}$ Dept. of Physics and Astronomy, University of Canterbury, Private Bag 4800, Christchurch, New Zealand \\
% ...
$^{21}$ Niels Bohr Institute, University of Copenhagen, DK-2100 Copenhagen, Denmark \\
$^{22}$ Dept. of Physics, TU Dortmund University, D-44221 Dortmund, Germany \\
% ...
$^{35}$ Dept. of Physics and Astronomy, University of Kansas, Lawrence, KS 66045, USA \\
% ...
$^{39}$ Dept. of Physics and Wisconsin IceCube Particle Astrophysics Center, University of Wisconsin{\textemdash}Madison, Madison, WI 53706, USA \\
% ...
$^{41}$ Department of Physics, Marquette University, Milwaukee, WI 53201, USA \\
% ...
$^{43}$ Bartol Research Institute and Dept. of Physics and Astronomy, University of Delaware, Newark, DE 19716, USA \\
% ...
$^{63}$ Deutsches Elektronen-Synchrotron DESY, Platanenallee 6, D-15738 Zeuthen, Germany \\
$^{\rm a}$ also at Institute of Physics, Sachivalaya Marg, Sainik School Post, Bhubaneswar 751005, India \\
% ...

\subsection*{Acknowledgments}

\noindent
The authors gratefully acknowledge the support from the following agencies and institutions:
USA {\textendash} U.S. National Science Foundation-Office of Polar Programs,
U.S. National Science Foundation-Physics Division,
U.S. National Science Foundation-EPSCoR,
U.S. National Science Foundation-Office of Advanced Cyberinfrastructure,
Wisconsin Alumni Research Foundation,
Center for High Throughput Computing (CHTC) at the University of Wisconsin{\textendash}Madison,
Open Science Grid (OSG),
Partnership to Advance Throughput Computing (PATh),
Advanced Cyberinfrastructure Coordination Ecosystem: Services {\&} Support (ACCESS),
Frontera and Ranch computing project at the Texas Advanced Computing Center,
U.S. Department of Energy-National Energy Research Scientific Computing Center,
Particle astrophysics research computing center at the University of Maryland,
Institute for Cyber-Enabled Research at Michigan State University,
Astroparticle physics computational facility at Marquette University,
NVIDIA Corporation,
and Google Cloud Platform;
Belgium {\textendash} Funds for Scientific Research (FRS-FNRS and FWO),
FWO Odysseus and Big Science programmes,
and Belgian Federal Science Policy Office (Belspo);
Germany {\textendash} Bundesministerium f{\"u}r Forschung, Technologie und Raumfahrt (BMFTR),
Deutsche Forschungsgemeinschaft (DFG),
Helmholtz Alliance for Astroparticle Physics (HAP),
Initiative and Networking Fund of the Helmholtz Association,
Deutsches Elektronen Synchrotron (DESY),
and High Performance Computing cluster of the RWTH Aachen;
Sweden {\textendash} Swedish Research Council,
Swedish Polar Research Secretariat,
Swedish National Infrastructure for Computing (SNIC),
and Knut and Alice Wallenberg Foundation;
European Union {\textendash} EGI Advanced Computing for research;
Australia {\textendash} Australian Research Council;
Canada {\textendash} Natural Sciences and Engineering Research Council of Canada,
Calcul Qu{\'e}bec, Compute Ontario, Canada Foundation for Innovation, WestGrid, and Digital Research Alliance of Canada;
Denmark {\textendash} Villum Fonden, Carlsberg Foundation, and European Commission;
New Zealand {\textendash} Marsden Fund;
Japan {\textendash} Japan Society for Promotion of Science (JSPS)
and Institute for Global Prominent Research (IGPR) of Chiba University;
Korea {\textendash} National Research Foundation of Korea (NRF);
Switzerland {\textendash} Swiss National Science Foundation (SNSF).\\
The authors gratefully acknowledge the computing time provided on the high-performance computer HoreKa by the National High-Performance Computing Center at KIT (NHR@KIT). This center is jointly supported by the Federal Ministry of Education and Research and the Ministry of Science, Research and the Arts of Baden-W\"urttemberg, as part of the National High-Performance Computing (NHR) joint funding program (https://www.nhr-verein.de/en/our-partners). HoreKa is partly funded by the German Research Foundation (DFG).\\
This project has received funding from the European Research Council (ERC) under the European Union’s Horizon 2020 research and innovation programme.

\end{document}